\documentclass{article}

\usepackage{arxiv}

\usepackage[utf8]{inputenc} 
\usepackage[T1]{fontenc}    
\usepackage{hyperref}       
\usepackage{url}            
\usepackage{booktabs}       
\usepackage{amsfonts}       
\usepackage{nicefrac}       
\usepackage{microtype}      
\usepackage{lipsum}		
\usepackage{graphicx}
\usepackage[numbers,square,sort&compress]{natbib}
\usepackage{doi}
\usepackage{amsmath}
\usepackage{tabularx}

\usepackage{textgreek}

\usepackage{textcomp}
\DeclareUnicodeCharacter{2212}{\textminus}

\newcommand{\citety}[1]{\citeauthor{#1} (\citeyear{#1})~\citep{#1}}

\title{Predicted state-to-state collisional rate coefficients from neural networks}

\author{ \href{https://orcid.org/0000-0000-0000-0000}{\includegraphics[scale=0.06]{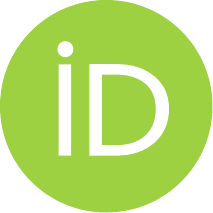}\hspace{1mm}Lis\'an David Cabrera-Gonz\'alez}\\
	Departamento de Física, Facultad de Ciencias, Universidad de Chile, \\
		Av. Las Palmeras 3425, 7800003 Ñuñoa, Santiago\\
	\texttt{lisan.cabrera@uchile.cl} \\
	\And
	\href{https://orcid.org/0000-0000-0000-0000}{\includegraphics[scale=0.06]{orcid.pdf}\hspace{1mm}Carlos Cárdenas} \\
	Departamento de Física, Facultad de Ciencias, Universidad de Chile, \\
		Av. Las Palmeras 3425, 7800003 Ñuñoa, Santiago \\
	Universidad Nacional de Colombia - Sede Medell\'in, Facultad de Minas, \\
    Departamento de Materiales y Minerales. Carrera 80 \# 65-223, \\
    N\'ucleo Robledo, Medell\'in, Colombia.\\
	\texttt{cardena@uchile.cl} \\
    \And
	\href{https://orcid.org/0000-0000-0000-0000}{\includegraphics[scale=0.06]{orcid.pdf}\hspace{1mm}Otoniel Denis-Alpizar} \\
	Departamento de Física, Facultad de Ciencias, Universidad de Chile, \\
		Av. Las Palmeras 3425, 7800003 Ñuñoa, Santiago \\
	\texttt{otoniel.denis@uchile.cl} \\
}

\begin{document}

    \maketitle
    
	
	
	\date{}
	
	
\begin{abstract}
	{Molecular spectra observed in the interstellar medium are commonly
analyzed using non-local thermodynamic equilibrium models, which
requires accurate state-to-state collisional rate coefficients. 
Usually, close-coupling calculations
become prohibitively expensive for transitions from highly excited rotational
states, therefore the available datasets are often truncated at low initial
states and temperatures; and since each new system also requires an accurate
\textit{ab initio} potential energy surface, a large number of detected molecules has no
collisional data available.}
	{We employ two machine learning strategies to
predict rotational state-to-state rate coefficients 
 of 
molecules included in the train set beyond the range of initial states
and temperatures computed so far, and predicting those of a molecule absent
from the train and validation sets.}
    { Feed-forward neural networks are trained on
state-to-state de-excitation rate coefficients of 24 linear molecules
colliding with para-H$_2$($j=0$), extracted from the BASECOL database. Scheme 1
includes all systems in the train+validation set, with boundaries in initial
state (70\%, 80\%, and 90\% of the initial states) and in temperature (90\% of the temperatures) defining the split
between training and prediction. Scheme 2 is trained on 22 systems
described only by transition and molecular properties, and
predicts two excluded systems, CS and CF$^+$, whose astrophysical impact
is explored with non-LTE radiative transfer calculations.}
    {For the 90\% boundary, Scheme 1 predicts 99.4\% of the
test transitions within a factor of 2, with an RMSE of $\log_{10}(k)$ of 0.07, and no prediction deviates from the reference value by more than a factor of 4.7.  This model was used
to complete the datasets of 18 systems up to 500~K and initial-state
energies up to 696~cm$^{-1}$. Scheme 2
reproduces the rate coefficients of two excluded systems within a
factor of 3 for more than 92\% of their transitions.}
    { Both schemes provide near-instantaneous complements
to close-coupling calculations: Scheme 1 extends the range of a molecule already included in the training set with high accuracy, while Scheme 2 gives a reasonable first estimate for a molecule that has not been studied. }
\end{abstract}

	

	\section{Introduction}
	Non-local thermodynamic equilibrium (non-LTE) models are crucial in analyzing regions of the interstellar medium (ISM) with low  densities. The Einstein coefficients and state-to-state inelastic scattering rate coefficients for the observed molecule in collisions with common interstellar colliders, particularly with H$_2$, are required to analyze observations using non-LTE models \citep{takComputerProgramFast2007}.  While the Einstein coefficients are usually known, this is different for the collision rate coefficients. \\
    
\noindent
    The absence of collisional data or the use of inaccurate rates can lead to significant errors in determining the physical and chemical conditions of the observed astrophysical regions \citep{denis2018new,bhat2022radiative,chandra2024maser}. The  community's demand for collisional rate coefficients has turned the study of molecular collisions into a highly active research field, with numerous groups around the world working on it.  In recent years, the volume of computed collisional data has increased, and specialized databases have been created to store and manage such data. The three primary databases are BASECOL \citep{dubernetm.l.BASECOL2023ScientificContent2024}, LAMDA \citep{van2020leiden}, and EMAA \citep{faure2025excitation}. \\

\noindent
   Despite the progress made in the field, many molecules still need collisional data to be estimated. 
    The number of systems for which rate coefficients datasets  are available  is about one-third of the $\sim$350 molecules detected so far in the interstellar medium or in circumstellar shells 
    \footnote{\url{https://cdms.astro.uni-koeln.de/classic/molecules}, accessed August 2026.}.  However, this number of molecules does not consider isotopes, therefore, the amount of necessary collisional datasets is even larger.\\

    \noindent
    Accurate collisional rate coefficients are traditionally obtained from close-coupling quantum dynamics calculations, which become computationally expensive as the initial rotational state increases, since higher-energy states require a larger rotational basis set. As a result, rate coefficients for transitions from high initial states are, in general, the most expensive to compute, and for many systems only data for the first rotational states and low temperatures are available.  Furthermore, computing the rate coefficients of a given collisional system is even more demanding, since it additionally requires the development of an interaction potential energy surface (PES) from \textit{ab initio} calculations. \\

\noindent
Approximate treatments are routinely used to reduce the computational cost, but each has a limited domain of validity. The Infinite-Order Sudden (IOS) approximation is  most reliable at collision energies well above the relevant rotational energy spacings. Its accuracy can deteriorate substantially at low collision energies, where rotational structure and threshold effects become important. Decoupling schemes such as the coupled-states approximation are generally more accurate, but also become less reliable at low collision energies.  However, these methods still require the availability of a suitable PES.\\

\noindent
On the other hand, when no data with H$_2$ exist, the rate coefficients computed with He are commonly mass scaled, but this has been shown to be inaccurate in many cases, particularly for ions \citep{yazidi2014revised,cabrera2020relaxation,tonolo2024collisional}. Also, the use of the rates for a similar systems are employed.
Moreover, all these procedures require data already available for the target molecule, or at least for a closely related one.\\

\noindent
Machine learning (ML) applications have already had a substantial impact on this field. ML-fitted PESs,  mostly based on neural
networks (NN) and on kernel methods such as Gaussian process and kernel
ridge regression, reproduce \textit{ab initio} interaction energies from a
comparatively small number of computed points, although the choice
of the sampling remains a non-trivial problem
\citep{krems2019bayesian,miksch2021strategies,liu2022full,kushwaha2026pes2mp}.\\

    \noindent
    The use of ML to determine collisional rate coefficients is still limited, and most of the existing work addresses vibrational rather
than rotational energy transfer, at temperatures well above those
relevant to the ISM. Complete datasets of vibrational state-to-state
rate coefficients have been generated by combining mixed
quantum-classical (MQC) dynamics with ML completion for
N$_2$+H$_2$ \citep{hong2023inelastic}, N$_2$+N$_2$ \citep{hong2023improved} and
CO+CO \citep{hong2024quantum}. More recently,
\citety{pezzellaVibrationalRelaxationRate2026} trained a NN on MQC rate coefficients for
CS+He in order to extend the dataset up to $v=45$ and 5000~K.\\

\noindent
Recently, \citety{bossionMachineLearningPrediction2024} used a multilayer
perceptron to predict accurate state-to-state rate constants for
H+H$_2$ from a large set of low-accuracy quasi-classical trajectories (QCT) values supplied as an
input feature, together with a small subset of accurate time-independent quantum mechanical (TIQM)
results. For SiO--H$_2$, \citety{wang2025multilayer} trained a multilayer
perceptron on the difference between coupled-channel and
decoupling-approximation results, substantially reducing the error
of the approximate calculations. In a related approach, \citety{mihalik2025accurate} combined quantum scattering
calculations with probabilistic ML models to obtain rate coefficients for
a large number of state-to-state transitions in molecule--molecule
collisions, with the model trained on rates 
computed for the same system. For rotational excitation
specifically, \citety{neufeld_rate_2010} pioneered the use of an NN to estimate rate coefficients of asymmetric top
molecules, reaching a median accuracy within a factor of $\sim$2 and
suggesting that such networks could be used to extrapolate existing
datasets towards high-lying transitions. ML have been mainly used to fill gaps of  rate coefficients in  systems for which scattering calculation have been already performed, or to improve the accuracy of calculations. However, to the best of our knowledge, no one has attempted to predict rate coefficients from simple molecular properties of the colliders, without resorting on partial information of the PES or scattering calculation of the target colliders. \\

\noindent
In this work we train feed-forward neural networks on the state-to-state
de-excitation rate coefficients of 24 linear molecules colliding with
para-H$_2$($j=0$). Scheme 1 extrapolates the rate
coefficients of a system already included in the training set towards the
higher initial states and temperatures. The model of this scheme is used to complete the rates of 18 systems for initial state energies up to 696 cm$^{-1}$ and temperatures up to 500K. The rates dataset of the remaining 6 systems already exist within these initial state energy and temperature. On the other hand, Scheme 2 predicts the rate coefficients of  molecules from its rotational spectra and molecular
properties alone. Information of the tests systems, CS and CF$^+$, was entirely excluded
from training.
The paper is organized as follows. Section 2 describes
the dataset and the two ML schemes, Section 3 presents our results
and their impact on a simple astrophysical
test, and Section 4 summarises our work.\\
	
	\section{Methodology}

		The rotational state-to-state collisional rate coefficients are obtained from the state-to-state cross sections $\sigma_{j \rightarrow j^{\prime}}\left(E_c\right)$ as
	\begin{equation}
		k_{j \rightarrow j^{\prime}}(T)=\sqrt{\frac{8}{\pi \mu k_B^{3} T^3}} \int_{0}^{\infty}\sigma_{j \rightarrow j^{\prime}}\left(E_c\right) E_c e^{-\frac{E_c}{k_B T}} \mathrm{d} E_c \, ,
		\label{rates}
	\end{equation}
	where $E_c$ is the collision energy, $T$ is the kinetic temperature, $\mu$ is the reduced mass of the collisional system, $k_B$ is the Boltzmann constant, and $j$ and $j^{\prime}$ are the rotational quantum numbers of the initial and final states, respectively. To first approximation, the energy of the rotational levels of a rigid linear molecule is given by
	\begin{equation}
		E(j) = B_e\, j(j+1) \, ,
		\label{e_be}
	\end{equation}
	where $B_e$ is the rotational constant of the molecule. Since the target of this work is a family of linear-molecule in collision with  \textit{para}-H$_2$($j=0$) the ML input features must capture both system-specific properties (e.g., mass, rotational constant) and transition-specific properties (e.g., the energy and quantum number of the initial and final states), together with the kinetic temperature.\\
    
    \noindent
	The de-excitation rate coefficients for 24 molecules were extracted from the BASECOL database \citep{dubernetm.l.BASECOL2023ScientificContent2024}, together with the temperature, and the quantum numbers and energies of the initial and final states of each transition. 
    Details of each dataset used in this work are shown in Table 3 of the Supplementary Information, hereafter SI.
    Molecular dipole moment, mass, rotational constant, ionization energy, and parallel and perpendicular polarizabilities were taken from the literature (see Table 1 of the SI). Because the rate coefficients $k$ span from $1.20\cdot10^{-32}$ (\textit{para}-H$_2$) to $5.82\cdot10^{-10}\,\mathrm{cm^3\,s^{-1}}$ (CN$^-$), the negative base-10 logarithm, $-\log_{10}(k)$ (ranging from 9.24 to 31.92), was used as the ML label instead \citep{bossionMachineLearningPrediction2024,neufeld_rate_2010}. \\
	
	\noindent
	The data was split into train, validation, and test sets. The train set  optimizes the NN parameters (weights and biases), the validation set optimizes  the hyperparameters, and the test set is reserved to evaluate the final model. All input features and output values were standardized using the mean and standard deviation of the train set only (Eqs. 1-3 of the SI), prior to training. The Pytorch \citep{NEURIPS2019_bdbca288} package was used to train the NN models. Along with Pytorch, the \texttt{hyperopt}\citep{bergstra2013hyperopt} Python package was employed to optimize the hyperparameters that determine the NN model for each scheme independently using a Bayesian optimization algorithm. This hyperparameter optimization was carried out to minimize the loss of the validation dataset using an early stopping criterion. The optimized hyperparameters were the activation function, the number of neurons of each hidden layer, the value of the alpha dropout regularization, and the learning rate (see Section 2 of the SI for more details).  \\
    
    \noindent
	Only rate coefficients for $T \ge 5$~K were kept. The data were restricted
to $T \le 500$~K, except for CO ($\le 3000$~K), HD ($\le 10\,000$~K), and
ortho/para-H$_2$ ($\le 6000$~K), for which the BASECOL datasets extend to
much higher temperatures and were retained in full during training in
order not to truncate these systems. All the results reported below are
restricted to $T \le 500$~K. 
    Since the rotational transitions of highest astrochemical interest are those with the largest rate coefficients, training and validation were restricted to $-\log_{10}(k) < 16$, while the test set was further restricted to predicted values of $-\log_{10}(k) < 15$ (to stay away from this boundary) and to initial-state energies below 696~cm$^{-1}$ (the energy levels generally retained in non-LTE the analysis, $E/k_B\sim 1000 K$, \citety{schoeierAtomicMolecularDatabase2005}).
	
	\subsection{Scheme 1. Predictions on molecules included in the train set}
	
    In Scheme 1, each transition is characterized by the features vector $\overset{\rightarrow}{x} = \left(\Delta j, \Delta E, T, m, B_e, \mu\right)$, combining the system properties (mass $m$, rotational constant $B_e$, dipole moment $\mu$) with transition properties (the difference in rotational quantum number $\Delta j$, the difference in state energy $\Delta E$, and the temperature $T$), with the output label as $y = -\log_{10}(k)$. These input features are of the same nature as those used by \citety{neufeld_rate_2010}. For each system, de-excitation rates from the lowest-lying initial states are used for training and validation, while rates from the highest-lying initial states are reserved to test the extrapolation performance of the trained model.\\

\begin{figure}
		\centering
		\includegraphics[width=0.5\linewidth]{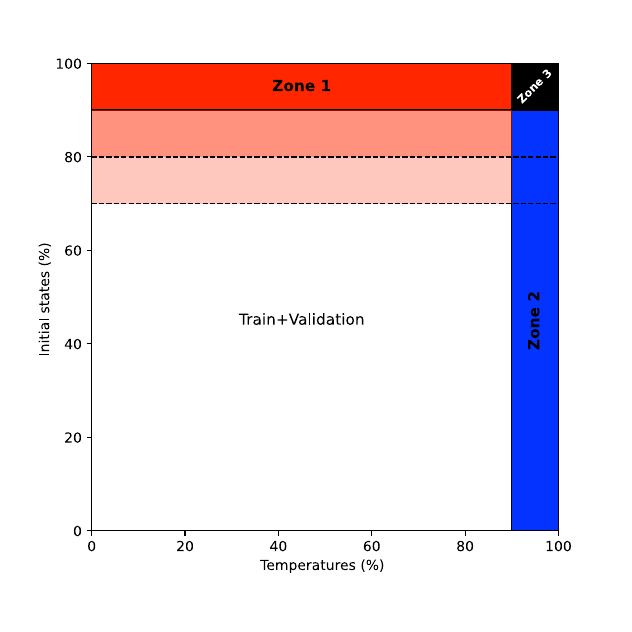}
		\caption{Graphical representation of the train+validation datasets employed to predict the test dataset split in 3 zones delimited by percentages of the initial states and temperatures of each system in Scheme 1.}
		\label{fig_data_1}
	\end{figure}

    \noindent
 The train, validation and test sets are chosen as follows. 
For each system, data is sorted by the energy of the initial state and temperature. Then, the data is split in train+validation and test. 
Train+validation contains data bounded by temperatures less or equal to the 90\% of the maximum number,  and initial energies that are also smaller or equal than a fraction (either 70, 80 or 90\%) of its largest numbers. 
The complementary data corresponds to the test set, that is thus divided into three zones in Fig. \ref{fig_data_1}: Zone 1, in which prediction of rates as a function of initial state and temperature correspond to an extrapolation and interpolation problem within the same system, respectively; Zone 2, where the opposite occurs; and Zone 3, where the dependency on both variables is an extrapolation problem. 
Using  three boundaries for the initial energies included in the train+validation set (70, 80 or 90\%) 
changes the  fraction of the data in the train/test set, which is useful to assess how much data is needed to achieve a certain degree of accuracy of the predictions. 
We call the models trained from these sets boundary 70/80/90\%. 
These percentages differ from the fraction of rate coefficients in the train+validation set, because the number of transitions per initial state increases with its index. 
Table~\ref{tab_points} gives the number of data points in each set and zone, together with the resulting train+validation fraction.\\

\noindent
An example of data splitting could be useful: the data for HCN have 12 initial states ($j=1-12$) and 11 temperatures
($T=5-100$~K). 
For boundary\%~$=90$\%, the train+validation set retains the
first $\lfloor 12\times0.9\rfloor = 10$ initial states and  the first
$\lfloor 11\times0.9\rfloor = 9$ temperatures, that is, $j=1-10$ at
$T=5-60$~K. 
The test set then contains $j=11,12$ at $T=5-60$~K (Zone 1),
$j=1-10$ at $T=80,100$~K (Zone 2), and $j=11,12$ at $T=80,100$~K (Zone 3).\\

	\begin{table}[h!]
		\centering
		\caption{Number of data points in each set and test zone for the three initial-state boundaries.      }   
		\label{tab_points}
		\begin{tabular}{lccc}
			\toprule
			 Boundary (\% of initial-state) & 70 & 80 & 90 \\
			\midrule
			Train+validation (fraction of all data) & 55.64\% & 68.16\% & 78.77\% \\
			Train+validation (size) & 32108 & 40982 & 50087 \\
			Test (size) & 25603 & 19142 & 13499 \\
			Test Zone 1 (size)  & 18848 & 12364 & 6718 \\
			Test Zone 2 (size)  & 3990 & 4961 & 5807 \\
			Test Zone 3 (size)  & 2765 & 1817 & 974 \\
			\bottomrule
		\end{tabular}
	\end{table}

\noindent
    For each Boundary\%, the train+validation set was further split into 85\% train and 15\% validation. To quantify the effect of this random split, ten NNs were trained with independent 15\% random train-validation splittings (RS); the predicted value ($y^{(\mathrm{pred})}$) is reported as the mean $\pm$ standard deviation over the ten RS models. The normalized distributions of $y$ for the train+validation and test sets are shown in Fig. 1 of the SI. The hyperparameters of NN architecture  were optimized with  with \texttt{hyperopt} \citep{bergstra2013hyperopt} only for boundary\% $=90\%$ and the first RS, and this same architecture was reused for the remaining Boundary\%s and RS (further details in section 2 of the SI). \\

	\subsection{Scheme 2. Predictions on molecules excluded from the train set}
	
	\begin{figure}[h!]
		\centering
		\includegraphics[width=0.5\linewidth]{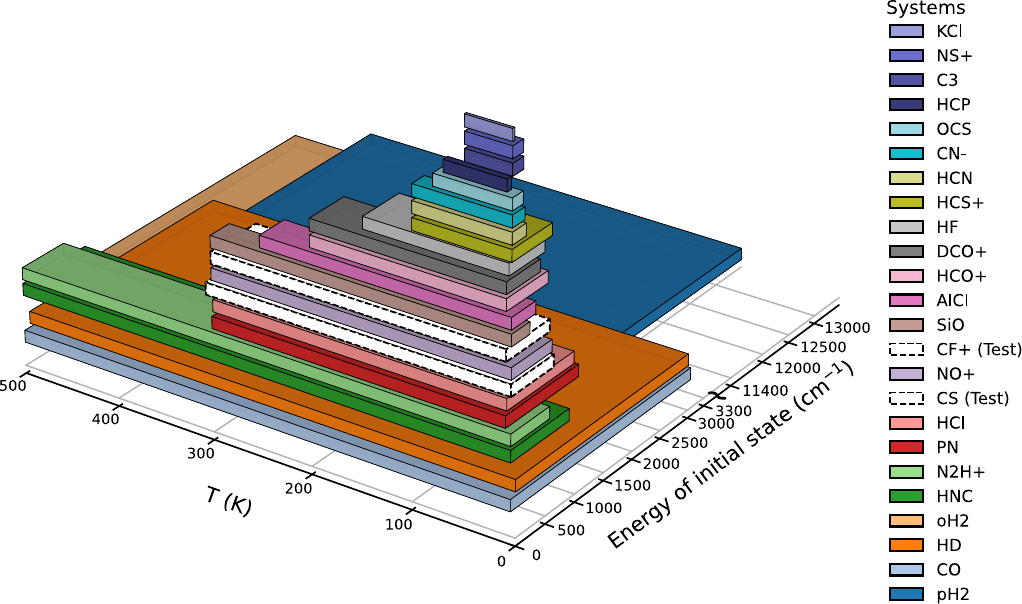}
		\caption{3D graphical representation of the train+validation (Color) and test (White) datasets for the prediction of rates of the two excluded systems (CS and CF+) by the NN model of Scheme 2. }
		\label{fig_data2}
	\end{figure}

\noindent
    In Scheme 2, the train+validation set consists of the transitions of all systems except the two test systems, CS and CF$^+$ (see Fig. \ref{fig_data2}). These two molecules were selected as the test dataset because one is a neutral molecule and the other is a charged  molecule. Furthermore, their rate coefficients and features distribution fall within the boundaries of the train+validation dataset distributions (see Tables 1-3 and Fig. 2 of the SI). Therefore, they are not seen at all during training. Within the train+validation dataset, ten non-random train-validation system splittings (SS) were performed, each selecting the full transitions of several  systems as the validation set, subject to two criteria: (1) the validation dataset size should be as close as possible to 15--20\% of the train+validation size, and (2) at least one charged and one neutral molecule must be included. The first SS consists of  HCS$^+$, SiO, and KCl as validation systems, and it was used to determine the NN architecture with \texttt{hyperopt}. All ten SS reuse this architecture, and the predicted rate coefficients are reported as the mean $\pm$ standard deviation over the ten SS. All the validation SS are shown in Table 4 of the SI. \\

    \noindent
    As Scheme 2 must generalize to systems absent from training, its input features rely on transitions and molecular properties. The transitions are described by the difference between the initial and final states energy ($\Delta E$), the initial and final rotational quantum numbers ($j_i,j_f$), and $T$; the systems are described by $B_e$, the first and second induction-interaction terms ($Ind_1,Ind_2$), a dispersion-interaction term ($Disp_1$), and the anisotropy of the  polarizability ($\Delta\alpha=\alpha_{||}-\alpha_\perp$), so that $\overset{\rightarrow}{x} = \left(T,\Delta E, j_i, j_f, B_e, Ind_{1}, Ind_{2}, Disp_{1}, \Delta \alpha \right)$. The induction and dispersion terms used are  \citep{Buckingham1978}:
	\begin{align}
		Ind_{1} &= -\frac{\pi \bar{\alpha}_B}{\left(4\pi \epsilon_0\right)^2}\left(\frac{q_A^2}{2R^4}\right) \, , \\
		Ind_{2} &= -\frac{\pi \bar{\alpha}_B}{\left(4\pi \epsilon_0\right)^2}\left(\frac{5\mu_A^2}{2R^6} \right) \, , \\
		Disp_{1} &= -\frac{\pi \bar{\alpha}_B}{\left(4\pi \epsilon_0\right)^2}\left(\frac{3IE_A\cdot IE_B \cdot \bar{\alpha}_A}{2 \left(IE_A+IE_B\right)R^6} \right) \, ,
	\end{align}
	where $\epsilon_0$ is the vacuum permittivity, the subscript $A$ denotes the target system and $B$ the collider \textit{para}-H$_2$($j=0$), treated as a sphere ($\Delta\alpha_B=0$, $q_B=0$, $\mu_B=0$, $\bar{\alpha}_B\neq0$, $IE_B\neq0$); the dipole moment, ionization energy, and polarizabilities are reported in Tables 1-2 of the SI. Note that we set the common  term $-\frac{\pi \bar{\alpha}_B}{\left(4\pi \epsilon_0\right)^2}$ to 1. The long-range distance $R$ was estimated from the mean polarizabilities following \citety{fedorovQuantumMechanicalRelationAtomic2018}:
	\begin{align}
		R &= 2.5 \cdot D(\bar{\alpha}_A,\bar{\alpha}_B) \, , \\
		D(\bar{\alpha}_A,\bar{\alpha}_B) &= 2\cdot2.54\cdot\left( \frac{\bar{\alpha}_A+\bar{\alpha}_B}{2} \right)^{\frac{1}{7}} \, ,
	\end{align}
	with $\bar{\alpha}=(\alpha_{||}+2\alpha_\perp)/3$ as the mean (isotropic) polarizability of each molecule.
	
	\section{Results and discussions}

    The performance of both schemes is quantified with the factor error ($FE$) between the rate coefficients predicted by the NN models and the true values from the BASECOL database on the corresponding test set (Eq. \ref{eq_fe}). $FE=1$ means the prediction equals the true value, while $FE=10$ means they differ by one order of magnitude. The other reported values for the schemes performance are the Mean Factor Error (MFE), the Root Mean Squared of the Factor Error (RMSFE), and the Maximum Factor Error (MaxFE). Such metrics allow us to get an insight into the mean error (MFE) of the rates, a mean error where the outliers have more weight (RMSFE), and outliers with the MaxFE. 

  \begin{equation}
		FE_i =  \frac{max(k^{(pred)}_{i},k^{(true)}_i)}{min(k^{(pred)}_{i},k^{(true)}_i)} \label{eq_fe} \\
	\end{equation}
    Where $k^{(pred)}_{i}$ and $k^{(true)}_i$ are the predicted and true rate coefficients for transition $i$, respectively. 
	
	\subsection{Scheme 1. Performance of predictions on molecules included in the training set}

	Scheme 1 predicts, for molecules included in the train set, the collisional rate coefficients of the transitions in the test set, which is made of the highest-lying initial states and highest temperatures. Therefore, these collisional rate coefficients (test set) are predicted using the transitions of the lower-lying initial states at the lower temperatures (train+validation) of the system itself, 
    and on the transitions of the other systems. \\

\begin{figure*}
		\centering
		\includegraphics[width=1\linewidth]{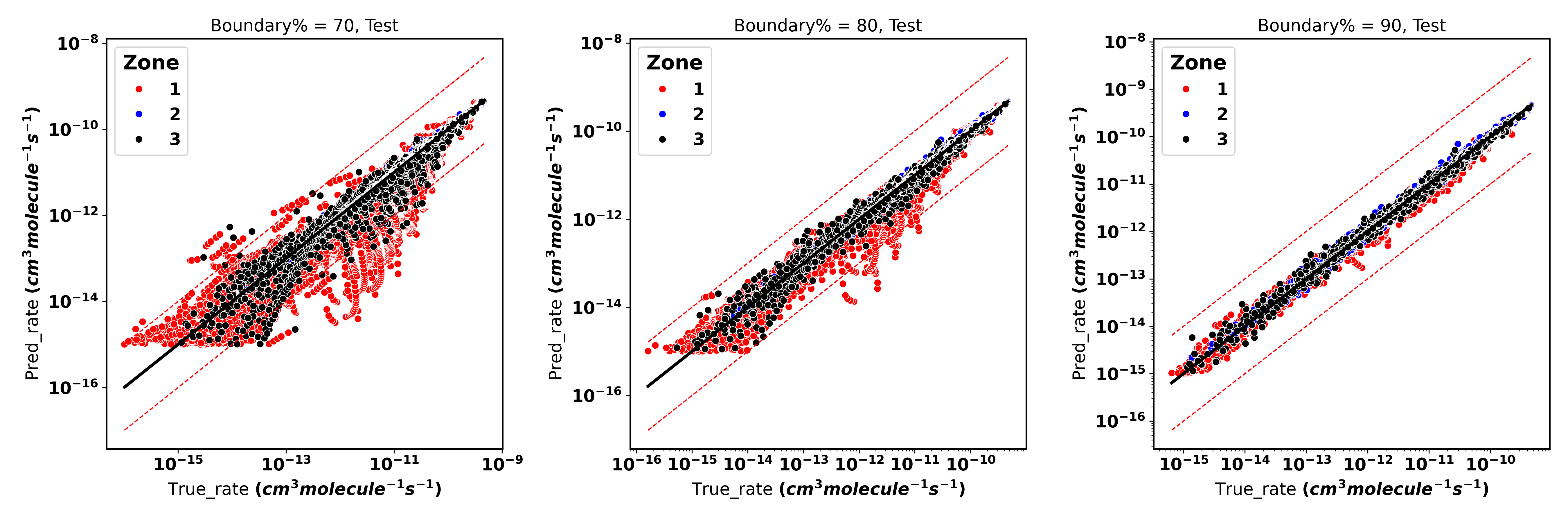}
		\caption{Comparison between the predicted and the true collisional rate coefficients as function of the zone and the boundary\%. The black solid line denotes the predicted = true line, whereas, the red dashed lines denote a difference of one order, that is, predicted = true */ 10.}
		\label{fig_predtrue}
	\end{figure*}

    \noindent
     Fig. \ref{fig_predtrue} shows the predicted and true rate coefficients of the test set. For Boundary\% = 70\%, the number of transitions exceeding one order of magnitude is 591, 552 of those belong to Zone 1 (552 represents 2.93\% of the points in Zone 1) and 39 of those belong to Zone 3 (1.41\% of points in Zone 3). Whereas, for Boundary\% = 80\%, the number that exceeds one order of magnitude  is 54, all of them belonging to Zone 1 (0.44\% of the points in Zone 1). The performance improves even further as the Boundary\% increases to 90\%, reaching a MaxFE of 4.7 (see Table \ref{tab_errors1}), that is, all the predictions are within a factor of 5 of the true values for all zones.\\

\begin{figure*}
		\centering
		\includegraphics[width=\linewidth]{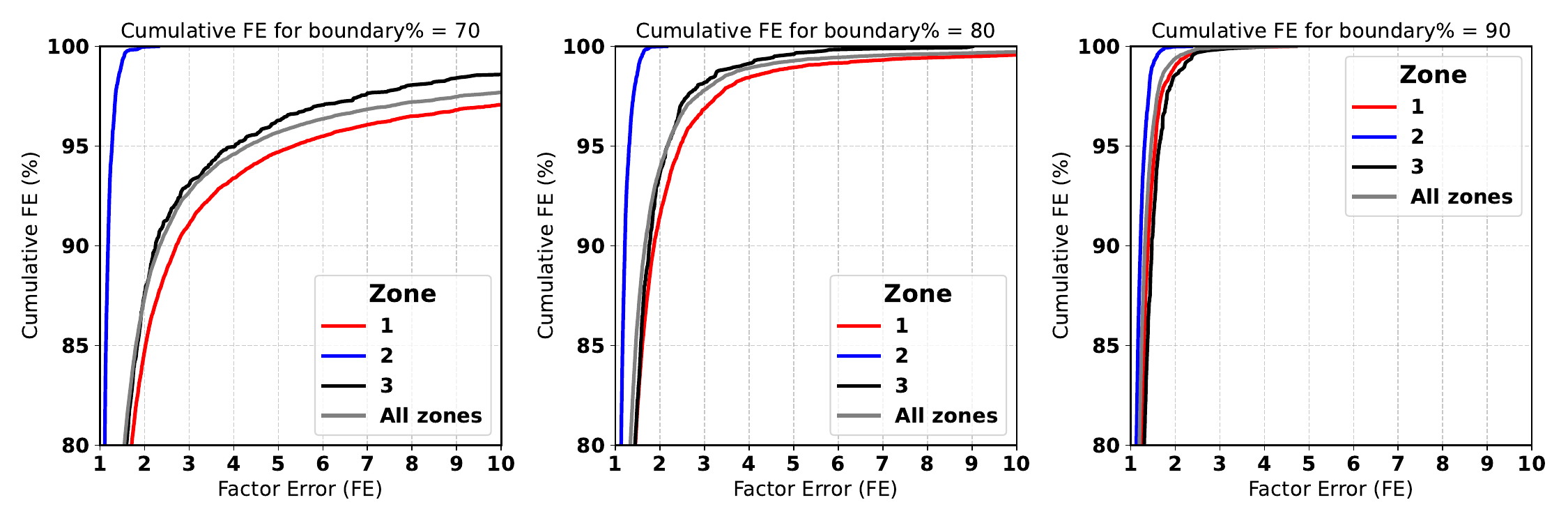}
		\caption{Cumulative error in (\%) of the FE (see Eq. \ref{eq_fe}) in the test dataset for each boundary\%.}
		\label{fig_scruve}
	\end{figure*}

	\noindent
    The cumulative-error distributions (see Fig. \ref{fig_scruve}) show that, for Boundary\% $=90\%$, 99.37\% of the test set is predicted with $FE<2$; this fraction decreases to 93.90\% and 87.40\% for Boundary\% $=80\%$ and $70\%$, respectively. Zone 2 consistently shows the best performance, reflecting the smooth dependence of the rate coefficients on temperature, whereas Zone 1 and Zone 3 behave similarly over most of the distribution. Zone 3 extrapolates along both axes (initial states and temperatures) at once, and its behaviour depends on the metric considered: for boundary\%~$=70\%$ and $80\%$ its tail is much better controlled than that of Zone 1 (RMSFE of 3.30 and 1.43 against 13.15 and 2.27, see Table \ref{tab_errors1}), whereas at boundary\%~$=90\%$, once the Zone 1 and Zone 2 errors become comparable, Zone 3 is marginally worse in the bulk of the distribution (MFE of 1.19 against 1.15) although its largest outlier remains smaller (MaxFE of 4.17 against 4.70).\\

    \noindent
A similar approach to the one presented in Scheme 1 was carried out by
\citety{neufeld_rate_2010}, who reports an RMSE of $\log_{10}(\mathrm{k})$ of
0.35 for ortho-H$_2$O in collision with para-H$_2(j=0)$. In our work we
obtain 0.30, 0.16, and 0.07 for the 70\%, 80\%, and 90\% boundaries,
respectively. Part of this difference reflects the nature of the target:
H$_2$O is an asymmetric top, whose level structure is considerably less
regular than that of the linear molecules considered here.\\

\noindent
 Additionally, the NN model for boundary\% = 90\% was employed to fill the missing collisional rate coefficients for all 18 systems from 5 K up to of 500 K and transitions with initial states up to 696 cm$^{-1}$, the datasets for CO, N$_2$H$^+$, HNC, para/ortho-H$_2$, and HD already cover such range. For rotational energies not included in the energy tables of BASECOL database, e.g. for rotational states larger than those reported, their energy was estimated by Eq. \ref{e_be}. These new collisional rate coefficients can be found as Supplementary Materials.
    
	\begin{table*}[h!]
		\centering
		\caption{Performance of the predicted collisional rate coefficients for the test dataset using the MFE, RMSFE, and MaxFE metrics for the several Boundary\% and Zones. The correlation between the predicted and true $-log_{10}(k)$ is represented by the R$^2$ value.}
		\label{tab_errors1}
		\begin{tabular}{lrrrrrrrrrrr}
			\toprule
			Metric & \multicolumn{3}{c}{Boundary\% = 70\%} & & \multicolumn{3}{c}{Boundary\% = 80\%} & & \multicolumn{3}{c}{Boundary\% = 90\%} \\
			\midrule
			R$^2$ & \multicolumn{3}{c}{0.949} & & \multicolumn{3}{c}{0.985} & & \multicolumn{3}{c}{0.996} \\
			MFE & \multicolumn{3}{c}{2.44} & & \multicolumn{3}{c}{1.33} & & \multicolumn{3}{c}{1.13} \\
			RMSFE & \multicolumn{3}{c}{11.35} & & \multicolumn{3}{c}{1.96} & & \multicolumn{3}{c}{1.14} \\
			MaxFE &\multicolumn{3}{c}{543.04} & & \multicolumn{3}{c}{80.87} & & \multicolumn{3}{c}{4.70} \\
			\midrule
			Metric  &  Zone 1 & Zone 2 & Zone 3 & & Zone 1 & Zone 2 & Zone 3  & & Zone 1 & Zone 2 & Zone 3 \\
			\midrule
			MFE & 2.83 & 1.07 & 1.75 & & 1.42 & 1.08 & 1.32 & & 1.15 & 1.08 & 1.19 \\
			RMSFE & 13.15 & 1.08 & 3.30 & & 2.27 & 1.09 & 1.43 & & 1.17 & 1.09 & 1.22 \\
			MaxFE & 543.04 & 2.30 & 66.01 & & 80.87 & 2.15 & 9.01 & & 4.70 & 2.36 & 4.17 \\
			\bottomrule
		\end{tabular}
	\end{table*}
	
	\subsection{Scheme 2. Performance of predictions on molecules excluded from the training set}

Scheme 2 trains ten NN models on the 22 systems included in the training set
to predict the rate coefficients of the two excluded systems, CS and CF$^+$
(see Fig.~\ref{fig_data2}). This is more challenging than Scheme 1:
with only 22 systems available, the system-describing features ($B_e$,
$Ind_1$, $Ind_2$, $Disp_1$, $\Delta\alpha$) take just 22 distinct
combinations, and the model must therefore predict at feature values it has
never encountered. This is also why the validation sets are built from
complete systems rather than from a random subset of transitions, and the
validation error then measures generalization to unseen molecules, which is
the quantity of interest here.\\

\begin{figure}
		\centering
		\includegraphics[width=0.45\linewidth]{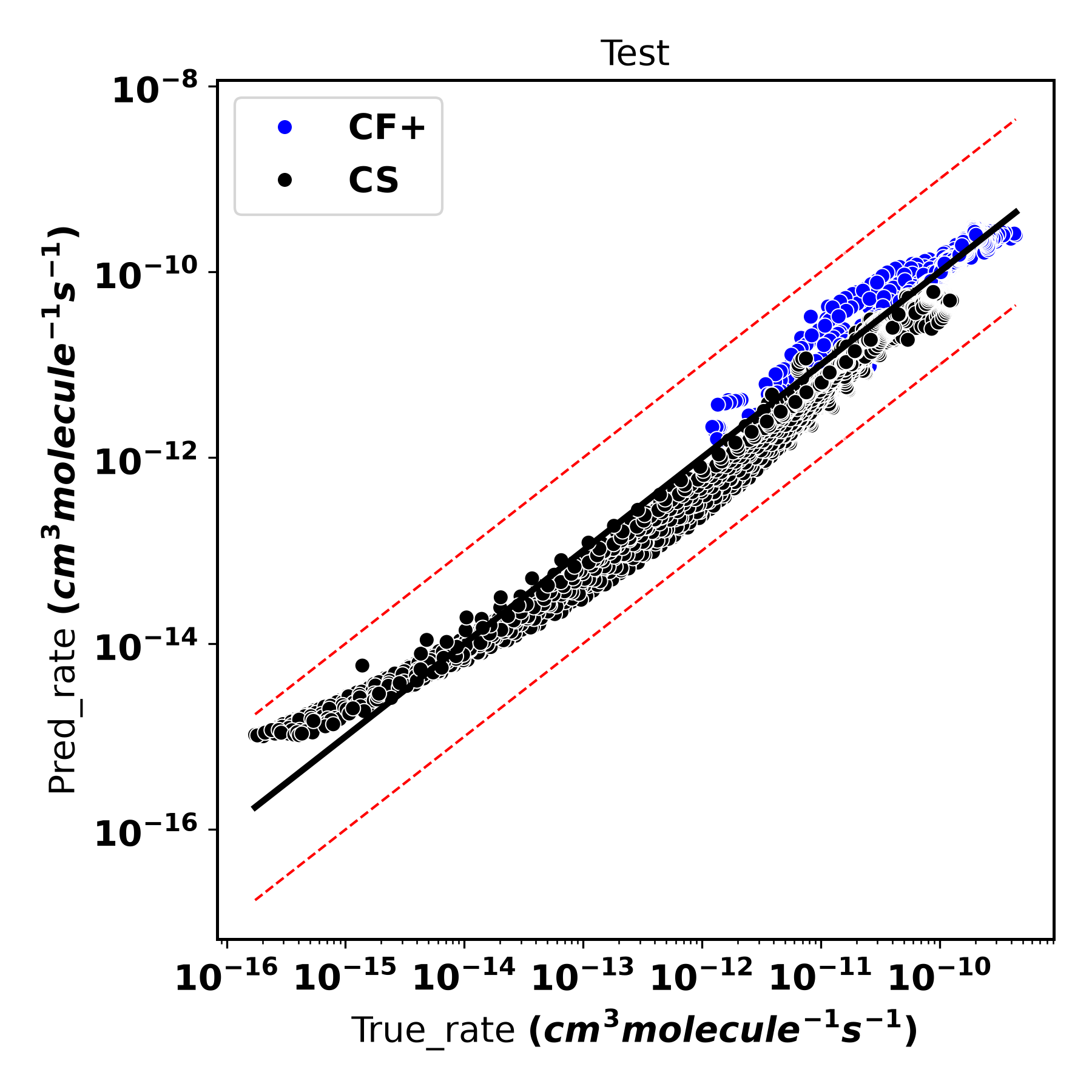}
		\caption{Comparison between the predicted and the true collisional rate coefficients for the test dataset systems CS and CF+. The black solid line denotes the predicted = true line, whereas, the red dashed lines denote a difference of 1 order of magnitude. The correlation $R^2$ value is presented in Table \ref{perf_sys}.}
		\label{fig_predtrue2}
	\end{figure}

\begin{figure}
		\centering
		\includegraphics[width=0.45\linewidth]{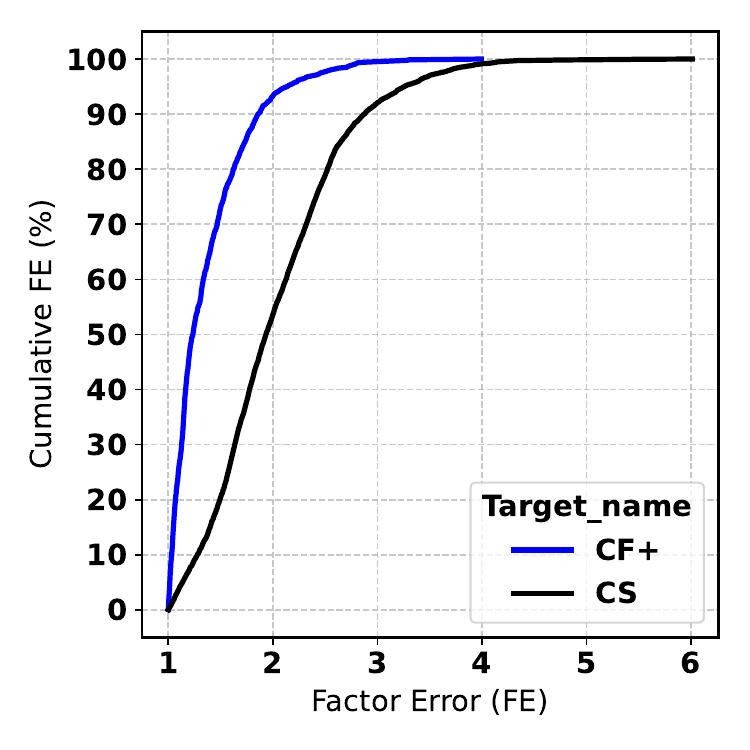}
		\caption{Cumulative error in (\%) of the FE (see Eq. \ref{eq_fe}) in the test dataset.}
		\label{fig_scruve2}
	\end{figure}

\noindent
	The predicted rate coefficients are more accurate for CF$^+$ than for CS (see Fig. \ref{fig_predtrue2}). The NN model predicts the CF$^+$ and CS rate coefficients within a factor of 2 for 93.16\% and 53.56\% of their transitions, respectively, and within a factor of 3 for 99.47\% and 92.05\% of their transitions, respectively (Fig. \ref{fig_scruve2}). Overall (see Table \ref{perf_sys}), the combined RMSFE is 1.99 and the MaxFE is 6.01, i.e., even the least accurate predictions remain within one order of magnitude of the true  value.   
 For CS, its rate coefficients span a range roughly four orders of magnitude larger
than those of CF$^+$, so a much larger fraction of its transitions lies in
the small-rate regime where the training data are the sparsest.  The better performance obtained for CF$^+$ should
therefore be read with this caveat in mind. \\

\noindent
The Relative Importance (RI) \citep{oldenIlluminatingBlackBox2002} of each input variable in scheme 2 is shown in Fig. 6 of the SI. Note that the RIs were determined with only the 10 SS values. With such note in mind, all the transition input variables ($\Delta E, j_i,j_f$) carry high RI, standing out $\Delta E$ and $j_f$, whereas, the RI of the temperature is not that high mostly due to the smooth behavior of the collisional rate coefficients with the temperature, as seen in Figs. 4-5 of the SI. On the other hand, note that among the input variables associated to molecular properties, the most relevant one is $\Delta \alpha$, which carry information of the anisotropy of the electron density of the target system. \\

    \noindent
    Since prediction accuracy is quantified via the factor of error with respect to the true reference values (FE), a metric that estimates accuracy without access to the true values is highly desirable for practical use of this NN model. As a first approach, we estimate prediction accuracy from the standard deviation of the 10 SS. Because the rates span many orders of magnitude, we use the mean coefficient of variation \citep{arachchige_robust_2022},
    \begin{equation}
        MCV = \frac{1}{N}\sum_{i=1}^{N}\frac{std_i}{mean_i},
    \end{equation}
    where $N = 1470$ and $4667$ for CF$^+$ and CS, respectively. 
    The MCV values are ordered consistently with the RMSFE of the two systems. To test
whether this holds at the level of individual transitions, we compared
$\log_{10}(std(k_i^{(pred)}))$ with $\log_{10}(|k_i^{true}-k_i^{pred}|)$
across all 6137 test points and found $R^2 = 0.871$ (see Fig. 7 of the SI).
  This correlation allows the prediction error to be estimated directly from the standard deviation of the predicted value, without requiring the true rate coefficient.

    \begin{table}[h!]
		\centering
		\caption{Performance of the predicted collisional rate coefficients for the test dataset using the MFE, RMSFE, and MaxFE metrics for the excluded CF$^+$ and CS systems. The correlation between the predicted and true $-log_{10}(k)$ seen in Fig. \ref{fig_predtrue2} is represented by the R$^2$ value. MCV stands for mean of the Coefficient of Variation \citep{arachchige_robust_2022}.}
		\begin{tabular}{lrrr}
			\toprule
			Metric & CF+ & CS & Total \\
			\midrule
			R$^2$ & 0.903 & 0.982 & 0.976 \\
			MFE & 1.37 & 2.04 & 1.88 \\
			RMSFE & 1.43 & 2.14 & 1.99 \\
			MaxFE & 3.99 & 6.01 & 6.01 \\
            MCV & 0.35 & 0.66 & 0.59 \\
			\bottomrule
		\end{tabular}
		\label{perf_sys}
	\end{table}
    
\subsection{Astrophysical implications}

To assess whether the Scheme 2 predictions are useful in practice, the quantum and predicted CS and CF$^+$ rate coefficients were used in the non-LTE radiative transfer code RADEX \citep{takComputerProgramFast2007}. We assumed a uniform sphere geometry and a background temperature of 2.73~K. Furthermore, we fixed the column density $N = 10^{12}$~cm$^{-2}$ and the line width $\Delta v = 1$~km~s$^{-1}$ for both molecules, which keeps the $j=1$-0, 2-1, and 3-2 lines optically thin ($|\tau| < 0.12$) over the full range of conditions explored.
The ratio $T_R(2\text{-}1)/T_R(1\text{-}0)$, where $T_R$ is the radiation temperature as defined in RADEX, is a common density diagnostic. We computed it as a function of the H$_2$ volume density, from $10^2$ to $10^8$~cm$^{-3}$, for kinetic temperatures of 10, 50, and 100~K. \\

\begin{figure}
		\centering
		\includegraphics[width=0.5\linewidth]{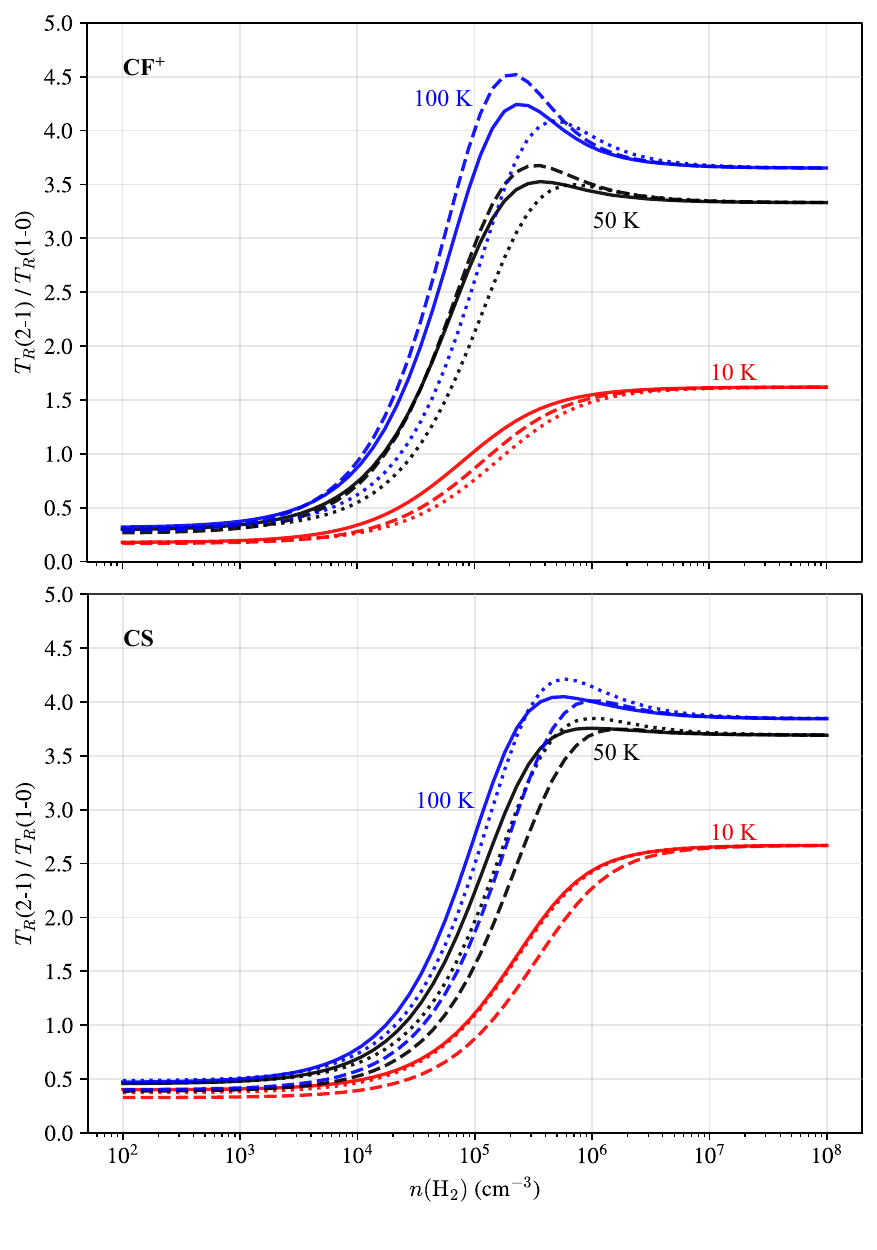}
		\caption{Ratio $T_R(2\text{-}1)/T_R(1\text{-}0)$ as a function of H$_2$ density for CS  and CF$^+$  at kinetic temperatures of 10, 50 and 100 K. Solid lines: using reference rates; dashed lines: using ML-predicted; dotted lines: using the He-scaled approximation.}
		\label{fig_ratio}
	\end{figure}

\noindent
Figure~\ref{fig_ratio} shows the $T_R(2\text{-}1)/T_R(1\text{-}0)$ ratio obtained with the ML-predicted rate coefficients (dashed lines) and with the reference close-coupling calculations (solid lines). For comparison, we also include the ratio obtained from the corresponding X--He close-coupling rate coefficients scaled by the factor of 1.4 recommended by \citety{schoeierAtomicMolecularDatabase2005}, taken from \citety{denis2018rotational} for CF$^+$ and \citety{lique2006rotational} for CS (dotted lines).
For $10^3 \le n({\rm H_2}) \le 10^6$~cm$^{-3}$, the Scheme~2 predictions for CF$^+$ deviate from the reference values by $8\%$ on average (up to $20\%$), compared with $18\%$ (up to $35\%$) for the He-scaled rates. 
This is not surprising as the mass-scaling approximation is expected to be less reliable for CF$^+$ because the long-range interaction of an ion with He and with H$_2$ differs. 
For CS, the He-scaled rates deviate by $5\%$ on average (up to $13\%$), while Scheme~2 deviates by $21\%$ on average (up to $35\%$). 
Above $\sim10^6$~cm$^{-3}$ the three sets converge as the lines
thermalize. Scheme~2 therefore provides a reasonable
first estimate whose accuracy depends on the system. Its practical advantage is that, unlike He scaling, it requires neither a prior He scattering dataset nor a PES for the target molecule.

	\section{Conclusions}

	We used ML strategies for increasing the size and the number of the datasets  available of collisional rates coefficients of molecules with astrophysical interest with para-H$_2$($j=0$). 
	Scheme 1 predicts the rate coefficients for a system included in the train and validation datasets toward higher initial states and temperatures, the regime that is most expensive to compute directly because it requires the largest rotational basis sets. Ten NN models, trained on rate coefficients from lower initial states and temperatures with identical architecture and hyperparameters, predict this regime with a mean factor error as low as 1.13 relative to the true BASECOL values at Boundary\% $=90\%$.The Scheme 1 model was used to complete the datasets of
 18 systems up to 500~K and initial-state energies of 696~cm$^{-1}$,
which are provided as Supplementary Materials. 

\noindent
Scheme 2 provides a rapid estimate of the rate coefficients of a system
that has not been seen by the model at all. Trained on 22 systems, each
described only by generic molecular properties, it predicts the rate
coefficients of two systems excluded from training, CS and CF$^+$, with a
maximum discrepancy of a factor of 6 relative to the true 
values. 
These predicted rate coefficients were included in RADEX, supporting their use as a first estimate of the collisional
properties of a newly identified linear molecule pending a full
quantum-dynamical calculation.\\

\noindent
Scheme 2 has two limitations that should be
kept in mind. The present models are restricted to linear molecules
colliding with para-H$_2(j=0)$, and the system-describing features take only 22 distinct combinations, so its extrapolation to
molecules far from the region of feature space covered here remains
untested. Yet,  a key message of this work is that predicting useful values of rate coefficient  from molecular properties of the colliders  seems to be a learnable problem for NN. Hence, as NN work better with large datasets, extending the approach to other colliders and type of targets, are natural continuations of this work.

	\section{Acknowledgements}
	This research has been supported by the following grant:
ANID/FONDECYT/EXPLORACIÓN/N.13250044. 

	%
    \bibliographystyle{unsrtnat} 
	\bibliography{biblio} 
	
\end{document}